\documentclass[aps,pre,twocolumn,showpacs,groupedaddress]{revtex4}

\usepackage{amssymb,amsmath}
\usepackage{graphicx}

\begin{document}

\title{Quantification of the differences between quenched and annealed 
averaging for RNA secondary structures}

\author{Tsunglin Liu and Ralf Bundschuh}

\affiliation{Department of Physics, Ohio State University, 191 W Woodruff Av., Columbus OH 43210-1117}

\begin{abstract}
The analytical study of disordered system is usually difficult due to the 
necessity to perform a quenched average over the disorder. Thus, 
one may resort to the easier annealed ensemble as an approximation to the 
quenched system. In the study of RNA secondary structures, we explicitly 
quantify the deviation of this approximation from the quenched 
ensemble by looking at the correlations between neighboring bases. This 
quantified deviation then allows us to propose a constrained annealed 
ensemble which predicts physical quantities much closer to the results of 
the quenched ensemble without becoming technically intractable.
\end{abstract}


\maketitle 

\section{Introduction}

Heteropolymer folding is of crucial significance in molecular biology.
It is the basis for the mechanism with which cells can produce three
dimensional building blocks out of the one-dimensional information
stored in their genome. Cells achieve this by forming (still
one-dimensional) polymers (proteins and RNA) by stringing together
different monomers with covalent bonds.  All monomers share a
compatible backbone but they have different side chains and occur in a
predefined order along the sequence. Physical interactions between
these monomers force the polymer to stably fold into a three
dimensional structure. This structure is crucial for the function of
the molecule; it is determined by the specific sequence of the
polymer~\cite{dill95,onuch97,garel97,shakh97}.

In addition to its biological relevance, heteropolymer folding is also
a very interesting problem of statistical mechanics~\cite{higgs00,higgs96,
pagna00,hart01,pagna01,bund02,krzak02,marin02,muller02,orland02,mukho03,
baiesi03,leoni03}. The competition
between the configurational entropy of the polymer, the overall
tendency of the monomers to stick to each other, the sequence
disorder, and the preference of folding toward a biologically active
native state, leads to a very rich thermodynamic phase diagram.  While
the same qualitative behavior is expected for proteins and RNA, we
will here concentrate on RNA since RNA folding is more amenable to
analytical and numerical approaches than protein folding. The relative
simplicity of the RNA folding problem compared to the protein folding
problem does not stem from the fact that RNA consists of only four
different bases versus the twenty amino acids the proteins are
composed of, but it comes from the simpler interaction rules: The
dominant interaction between the four bases A, U, G, and C of an RNA
molecule is Watson-Crick (G--C and A--U) pair formation, i.e., if two
bases have formed a pair they to first order do not take part in any
further interactions. Every amino acid of a protein on the contrary
interacts with all its spatial neighbors, i.e., with on the order of
ten other amino acids at a time.

From a statistical physics point of view, the possibility of a glass
phase at low temperatures driven by sequence disorder, is of special
interest in the heteropolymer folding problem~\cite{higgs96,pagna00,hart01,
pagna01,bund02,krzak02,marin02}. Unfortunately, even
for the case of RNA folding an analytic quantitative description of
the glass phase is still outstanding. Thus, quantitative studies have
to either rely on numerics or they have to use what is known as the
annealed average.  In the annealed average, the free energy of the
system is approximated as the logarithm of the ensemble averaged
partition function (instead of taking the ensemble average over the
logarithm of the partition function called the quenched
average). Physically, this approximation corresponds to treating the
sequence degrees of freedom as dynamical instead of frozen variables.
Thus, the annealed system represents a sequence ensemble that is
coupled to the structural ensemble by way of the interaction
energies. This sequence ensemble may be different from the original
sequence ensemble of uncorrelated random sequences over which the free
energy is supposed to be averaged. Due to these differences between
the annealed and the quenched sequence ensemble the annealed free
energy is only an approximation to the true (quenched) free energy of a 
disordered system. 

The purpose of this manuscript is to first quantify the differences between 
the annealed and the quenched sequence ensembles. Specifically, we will look 
at correlation between neighboring bases. We show that while this
correlation is strictly zero in the correct (quenched) sequence
ensemble, they are non-zero in the annealed sequence ensemble and
increase with decreasing temperature - up to complete correlation in
certain models of RNA folding. This clearly underlines and quantifies
the fundamental shortcomings of the annealed average in the RNA
folding problem at low temperatures.

Based on the quantified non-zero nearest neighbor correlations, we then 
try to diminish the differences between the annealed and quenched 
ensembles by forcing the annealed ensemble to present zero neighboring 
correlation. This constrained annealed ensemble behaves much more 
similar to the quenched ensemble than the annealed ensemble. Although 
the glass phase itself can not be identified using the constrained 
annealed ensemble which only partially corrects the overall non-random 
correlations, one can obtain thermodynamic quantities which are much 
closer to the quenched results than the annealed ones using this method. 

This paper is organized as follows: 
In Sec. II, we briefly review the RNA secondary structure and introduce the 
general RNA folding problem with sequence disorder. In Sec. III, we quantify 
the deviation of nearest neighbor correlations of the annealed ensemble. 
Finally, we improve the pure annealed ensemble by applying a constraint of 
random correlations in Sec. IV.

\section{RNA folding problem with sequence disorder}

\subsection{RNA secondary structures}

RNA is a single-stranded biopolymer of four different bases A, U, C, and G. 
The strand can fold back onto itself and form helices consisting of stacks 
of stable Watson-Crick pairs (A with U or G with C). This comparatively 
simple interaction scheme makes the RNA folding problem very amenable to 
theoretical approaches without losing the overall flavor of the general 
biopolymer folding problem~\cite{higgs00}.

An RNA secondary structure {\bf S} is characterized by its set of Watson-Crick 
base pairs $(i,j)$ where $i$ and $j$ denote the $i^{th}$ and $j^{th}$ base 
 of the RNA polymer respectively (conventionally $i<j$). Here, we follow many 
previous studies~\cite{higgs00,higgs96,pagna00,hart01,pagna01,bund02,
krzak02,marin02,muller02,orland02,mukho03,baiesi03,leoni03} and apply the 
reasonable approximation to exclude 
so-called pseudoknots~\cite{tinoco}, i.e., for two Watson-Crick pairs 
$(i,j)\in {\bf S}$ and $(k,l)\in {\bf S}$, configurations with $i< k < j < l$ 
are not allowed. This approximation is justified, because short pseudoknots 
do not contribute much to the overall energy and long pseudoknots are 
kinetically difficult to form. 

\subsection{Quenched averaging}

The properties of RNA folding, especially the possibility of a glass phase 
driven by the sequence disorder, have been a challenging problem from the 
statistical physics points of view. To understand the statistics of this 
disordered system, one first has to assign an energy $E(\chi,{\bf S})$ to 
every 
secondary structure {\bf S} for a given sequence $\chi$. This could, e.g., 
simply be the negative of the total number of Watson-Crick base pairs. This 
then allows us to calculate the partition function 
\begin{equation}
Z(\chi)=\sum_{{\bf S}} \cap(\chi,{\bf S}) e^{-E(\chi,{\bf S})/T} 
\end{equation}
for a given sequence $\chi$ where $\cap(\chi,{\bf S})$ is one when the 
secondary structure ${\bf S}$ is compatible with the sequence $\chi$ and zero 
otherwise. Finally, one has to calculate the quenched average
\begin{equation}
F_q = -k_B T \langle \ln Z(\chi) \rangle_{\chi}
\end{equation}
over all sequences $\chi$.

\subsection{Annealed averaging}

Unfortunately, the quenched free energy $F_q$ is very difficult to calculate. 
Thus, one 
can try to approximate the quenched free energy by the much easier computed 
annealed free energy, which treats the disordered sequences as dynamic 
variables. This annealed free energy is only a lower bound of the quenched 
free energy, 
\begin{equation}
F_a = -k_B T \ln \langle Z(\chi) \rangle_{\chi} < F_q. 
\end{equation}

It can be quite different from the quenched free energy since the 
annealed ensemble favors those 
sequences where more binding pairs are allowed. More importantly, 
physical quantities derived from this annealed free energy can be very 
different from their quenched counterparts as we will show explicitly 
in the following sections.  
To be specific, we will measure the correlation between neighboring bases 
which are known to vanish in the quenched case. 

\subsection{Energy models}

In this paper, we study the simplest model of disordered RNA sequences 
which contain only the two bases A and U. In assigning free energies to 
secondary 
structures, we neglect any loop entropies and focus on the base pairs 
alone. Besides, for most parts of this manuscript, we do not consider the 
minimal hairpin length constraint which requires the two bases of a binding 
pair to be separated by at least three bases in a real RNA molecule. Within 
these approximations we do consider two different energy models.

In the {\em binding energy model}, we simply assign an energy $\epsilon=-1$ to 
each AU (or UA) binding pair. We denote the corresponding Boltzmann factor by 
$q=e^{1/T}$. This model captures the main features of the energetics and is 
simple enough for analytical and numerical studies. 

We also study a somewhat more realistic energy model, namely the {\em stacking 
energy model}. In this model, we assign energies to the stacking of two base 
pairs rather than to individual base pairs. This stacking energy depends in 
reality on the identities of all four bases involved. We 
implement this effect by associating a Boltzmann factor $s_1$ with stackings 
of types ${AA \atop UU}$ and ${UU \atop AA}$ while associating a different 
Boltzmann factor $s_2$ 
with stackings of types ${AU \atop UA}$ and ${UA \atop AU}$. To be specific, 
we will choose 
these Boltzmann factors as $s_1=e^{2/T}$ and $s_2=e^{1/T}$ for the remainder 
of this communication.

The main reason to study the stacking energy model is that the simple 
binding energy model is known to be 
pathological without a glass phase at low temperature in the disordered 
sequence ensemble~\cite{pagna00,hart01,pagna01}. A simple reason is that 
whatever the 
sequence, each base A can always find another base U to pair with provided we 
have the same amount of bases A and U. Thus, sequences disorder does 
not cause frustration. In contrast, the energy distribution 
of the stacking energy model is greatly affected by sequences, and a structure 
in which all base pairs are stacked can in general not be found for every 
sequence. 
Thus, sequence disorder is expected to cause frustration, and a glass phase 
is expected in this energy model for low enough temperature.

\section{Nearest neighbor correlations of the annealed ensemble}

In this section, we calculate quantitatively how the nearest neighbor 
correlations in the annealed ensemble deviate from their true values in 
the random sequence ensemble. To this end, we have to calculate the annealed 
partition function for sequences with length N-1, which is defined as 
\begin{equation}
Z_a(N) = \frac{1}{2^{N-1}}\sum_{{\bf S}} \left( \sum_{\chi} \cap(\chi,{\bf S}) 
e^{-E(\chi,{\bf S})/T} 
\right).
\end{equation}

\begin{figure}[h]
\includegraphics[width=8cm]{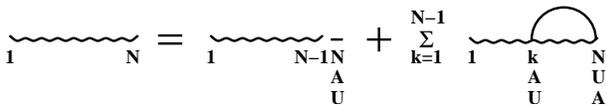}
\caption{Recursive relation exploring all possible secondary structures for 
a homogeneous sequence of length N. The wavy lines stands for contribution 
from all possible structures and sequences. The straight line stands for 
non-paired bases.}
\label{molten}
\end{figure}

For the binding energy model, this annealed partition function can be easily 
obtained via the recursive relation shown in Fig.~\ref{molten} along the lines 
of previous studies~\cite{degenn68,water78,zuker84,mccas90,bund02} but 
taking the sequences into account explicitly. The idea is 
to separate the two cases for the last base, which is either unbound or bound 
to a certain base k, and then relate the partition function to the shorter 
length one as 
\begin{eqnarray}
Z_a(N+1;q) &=& Z_a(N;q) \\ \nonumber 
&+& \frac{q}{2} \sum_{k=1}^{N-1} Z_a(k;q)Z_a(N-k;q).
\end{eqnarray}
With this relation, one can obtain an analytical 
formula for the annealed partition function in the large N limit by performing 
the z-transform, which is defined as 
\begin{equation}
\widehat{Z_a}(z;q)=\sum_{N=1}^{\infty} Z_a(N;q)z^{-N},
\end{equation}
on the recursive relation. After solving the resulting quadratic equation 
for $\widehat{Z_a}(z;q)$, we can 
obtain the partition function by doing the inverse z-transform, 
\begin{equation}
Z_a(N;q)=\frac{1}{2\pi i} \oint \widehat{Z_a}(z;q) z^{N-1} dz.
\end{equation}
This approach can be easily generalized to the stacking energy model.

In order to keep track of the correlations by the annealed ensemble, we 
assign an additional Boltzmann factor $L$ to all AA and UU neighbors 
within the sequence. The modified annealed partition function is then 
\begin{equation}
Z_a(N;q,L) = \frac{1}{2^{N-1}}\sum_{{\bf S}} \left( \sum_{\chi} 
\cap(\chi,{\bf S}) q^{n_q({\bf S})} L^{n_L({\chi})} \right),
\end{equation}
where $n_q({\bf S})$ is the number of binding pairs in a secondary 
structure ${\bf S}$, and $n_L({\chi})$ is the number of conjugate neighbors, 
i.e., AA and UU neighbors in the sequence.

The additional Boltzmann factor complicates the calculation of the 
partition function since different bases A and U contribute differently. 
However, we can still formulate recursive relations by noticing that the 
two end bases of a sequence piece determine the correlations with other 
pieces. Thus, we can separate a sequence into two cases where the end bases 
are either of the same type or not, and formulate the recursive relation for 
each case independently. The annealed partition function $Z_a(N;q,L)$ is then 
obtained via z-transform as before. Since the formation of the recursive 
relations is quite technical, we only address the result here, and defer 
the details to Appendix A.

From the partition function we can obtain the nearest neighbor 
correlations by looking 
at the deviation of the averaged fraction of AU (or UA) neighbors from the 
expected value 1/2 in the disordered sequence ensemble. This deviation $\delta$ is obtained by taking the derivative as
\begin{equation}
\delta=\frac{1}{2}-\frac{1}{N}L \partial_L \ln(Z_a(N;q,L))|_{L=1}.
\end{equation}

\subsection{Binding energy model}

\begin{figure}[h]
\includegraphics[width=8cm]{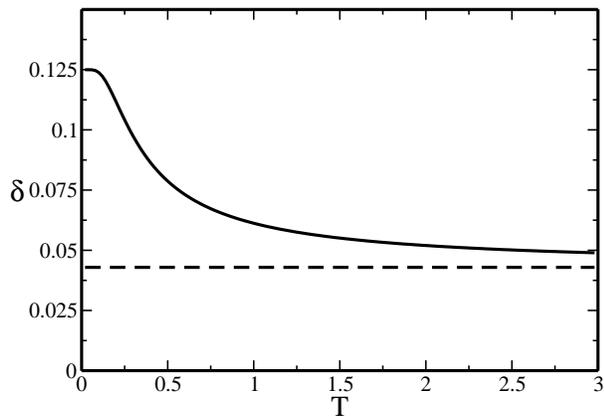}
\caption{Deviation of the fraction of AU (or UA) nearest neighbors. The 
deviation is plotted as a function of temperature 
in units of the binding energy for the binding energy model. Notice that 
the deviation moves further away from zero and stops at a fixed constant as 
temperature decreases. It also approaches a limit larger than zero at high 
temperature indicated by the dashed line.}
\label{frLq}
\end{figure}

Fig.~\ref{frLq} shows the neighbor correlations for the binding energy model. 
We find that the deviation moves further away from zero as temperature 
decreases. This is a direct result from the fact that 
at low temperature, the main contributions to the annealed partition function 
come from those sequences which allow a lot of binding pairs, unlike the 
quenched case where sequences are equally weighted. 

The exact way that the neighbor correlations are biased can be understood 
as follows. In this binding energy model, the only thing that biases the 
nearest neighbor correlations is the formation of minimal hairpins since 
they enforce the neighboring bases to be different, which are either AU or 
UA. Thus, the degree of bias is 
directly coupled to the fraction of smallest hairpins in a sequence. 

This assertion can be verified by studying the fraction of minimal hairpins. 
As an example, we study the zero temperature case where all the bases are 
expected to be paired. 
Among all possible pairing structures, we explicitly calculate the fraction 
of smallest hairpins (with the details shown in Appendix B). 
As a result, every fourth base is part of a minimal size hairpin. 
Thus, we have $1/4$ AU (or UA) nearest neighbors from these hairpins and 
another 
$1/2\times3/4=3/8$ from the rest of the bases since they do not show nearest 
neighbor correlation bias. The deviation of the fraction of AU (or UA) 
neighbors is then expected to be $5/8-1/2=1/8$, which 
matches exactly the zero temperature limit in Fig.~\ref{frLq}.
In this case, the sequence, as a dynamic variable, adjusts 
itself to all the binding pairs. 

Even in the high temperature limit, although all allowed sequences are equally 
weighted, there still exists a finite fraction of minimal size hairpins on 
average. As a result, the deviation of neighbor correlations approaches a 
constant larger than zero. 

The assertion that the deviation $\delta$ is coupled to the formation of 
minimal size hairpins is again verified as we additionally require 
all the hairpins being of length larger than one. In this case, the 
correlation between 
nearest neighbors becomes random at all temperatures. However, the second 
nearest neighbor correlations become non-trivial.

This simple binding energy model gives us a taste how the nearest neighbor 
correlations are coupled with the energy through the structure, i.e., the 
formation of minimal hairpins. This correlation is biased since the 
annealed ensemble puts more weight on lower energy sequences. 

\subsection{Stacking energy model}

Following the same approach, we check the same deviation 
as a function of temperature in the more realistic stacking energy model. 
Again, only the result is 
quoted here in Fig.~\ref{frLs12} (interested readers can check the detailed 
calculations in Appendix C).

\begin{figure}[h]
\includegraphics[width=8cm]{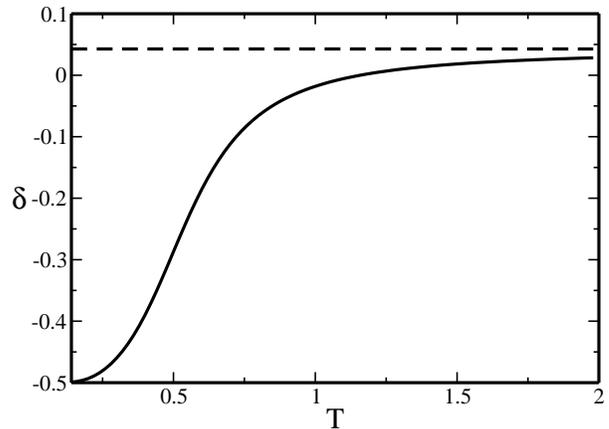}
\caption{Deviation of the fraction of AU (or UA) nearest neighbors 
for the energy model involving stacking energies. Unlike in the case of 
the binding energy model, the AU (or UA) neighbor correlations are 
completely biased at zero temperature in the stacking energy model. At 
high temperature, this deviation approaches the same limit as the binding 
energy model.}
\label{frLs12}
\end{figure}

Unlike the binding energy model, at zero temperature, the nearest neighbor 
correlations of the stacking energy model are completely biased. Almost 
no AU (or UA) neighbors can be found in this annealed system. This can 
be understood since at zero temperature, the only dominating structure is 
a long stem in which all stacking loops are of type $s_1$. Thus, the only 
two important sequences are the ones made of half consecutive A bases and 
the other half of U bases. 

\begin{figure}[h]
\includegraphics[width=8cm]{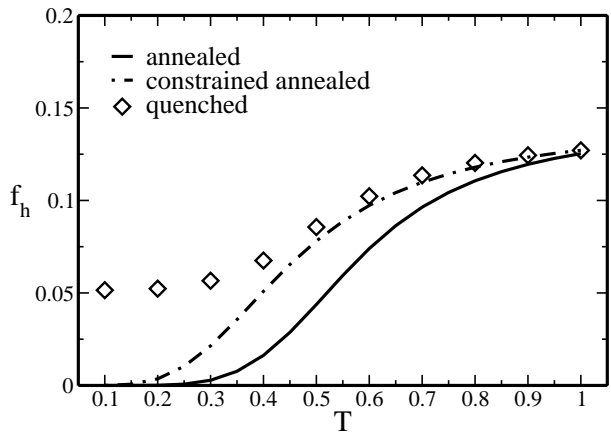}
\caption{Fraction of hairpins in the stacking energy model for three 
different ensembles.}
\label{frhs12}
\end{figure}

To verify this structure, we additionally introduce another Boltzmann factor 
$h$ for each hairpin loop formation. With this Boltzmann factor we can keep 
track of the fraction of hairpins $f_h$ in the annealed system by calculating 
\begin{equation}
f_h=\frac{1}{N}h \partial_h \ln(Z_a(N;s_1,s_2,h,L=1))|_{h=1}.
\end{equation}
From Fig.~\ref{frhs12}, 
we do see that the fraction of hairpins of this annealed system indeed goes 
to zero as temperature goes to zero, which is a feature of the long stem 
structure. 

At high temperature, however, the energy model does not matter since 
entropy dominates. Thus, the AU (or UA) fraction approaches the same limit 
as in the binding energy model. 

From this stacking energy model, we learn that the stronger the energy is 
coupled to the nearest neighbor correlations, the larger deviation in 
nearest neighbor correlations of the annealed system will be present at low 
temperature.
 
\section{Constrained annealing}

So far we have only observed the sequence correlations artificially 
introduced through the annealed ensemble. However, our approach can in fact 
be used to generate more realistic ensembles within the annealed framework. 
The idea is to force the nearest neighbor correlations to be random when 
performing the annealed average~\cite{morita64,orlandi02}. 

We simply enforce this random disorder constraint, i.e., the fraction of 
AU (or UA) 
neighbors being one half by setting the Boltzmann factor $L$, which controls 
the nearest neighbor correlations, to whatever value it needs to have for the 
correlations of the annealed ensemble to vanish. 

This constrained annealing turns out to predict 
thermodynamic quantities much closer to the quenched results. And it can 
be done immediately following our quantified deviations in disorder. 

\subsection{Binding energy model}

The constraint for the binding energy model is read as
\begin{equation}
\frac{1}{N}L \partial_L \ln(Z_a(N;q,L))|_{L=L_c} = \frac{1}{2}. 
\end{equation}
In this energy model, we expect the sequences with more AU (or UA) nearest 
neighbors to be suppressed since the annealed system favors those neighbors. 
As a result, $L_c$, which favors AA (or UU) neighbors, is expected to be 
larger than one in order to meet the constraint. Furthermore, $L_c$ should be 
larger at lower temperatures since the neighbor correlation is more biased 
at lower temperatures.

One important note is that the resulting free energy is only defined up to 
an additive constant, i.e., adding a constant background potential does 
not change the result at all. Thus, the absolute value of this constrained 
annealed free energy as well as the Boltzmann factor $L_c$ has no real 
meaning. For example, one could assign the Boltzmann factor $L$ 
to AU (or UA) neighbors instead of AA (or UU) neighbors. The resulting 
chemical potential would then change a sign and the free energy would differ 
by a constant amount. However, the thermodynamic quantities, 
which are calculated by taking derivatives of the constrained free energy, 
will not see this constant and are expected to be closer to the quenched 
result. 

To verify this assertion, we are going to compute the average fraction of 
binding pairs for 
the binding energy model via $q/N \partial_{q} \ln(Z_a(N;q,L))$ as a 
function of temperature. Then, we compare the cases of the annealed ($L=1$), 
the constrained annealed ($L=L_c$) and the quenched ensembles. 

\begin{figure}[h]
\includegraphics[width=8cm]{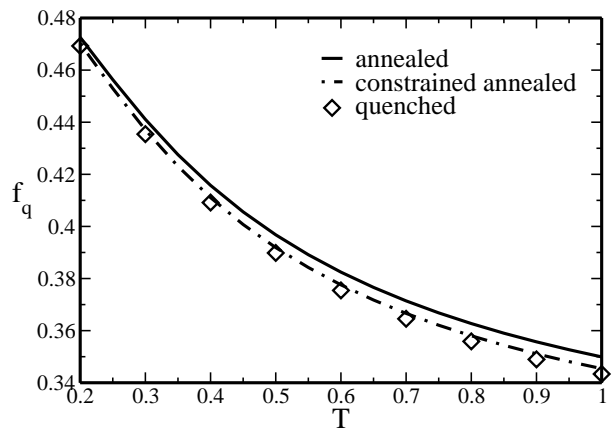}
\caption{Fraction of binding pairs in the binding energy model. The 
constraint of random nearest neighbors brings the annealed quantity closer 
to the quenched numerical estimate. The statistical errors of the quenched 
results are always smaller than the size of the symbol.}
\label{frq}
\end{figure}

As to the quenched result, we numerically calculate the partition function 
given random sequences of length 1280 and collect the data from 1000 
random sequences. In order to avoid the trivial finite size effects 
due to fluctuation of the fraction of A bases away from its expected value 
1/2, we only choose sequences that contain exactly 640 A's and 640 U's. 
The result is shown in Fig.~\ref{frq}. The statistical errors of the 
quenched results are always smaller than the size of the corresponding 
symbol, such that within the error bars the quenched results never overlap 
other curves. This condition holds for all other quenched results in this 
manuscript.

We see that the constrained annealed result is indeed very close to 
the quenched numerical estimate. However, all three results are rather close 
to each other anyway. The reason for these three cases being so close 
to each other is simply that under this energy model the system is not glassy, 
and every base is able to find another base for 
pairing in this binding energy model. Thus, at zero temperature, all the 
bases are paired in all three systems. The fact that the nearest neighbor 
correlations are not biased a lot can also be verified as we find that at 
T=0.1, $L_c$ to be just 1.59. Thus, the chemical potential introduced 
from the constraint is comparatively small and does not affect the result 
too much.

\subsection{stacking energy model}

The situation for the stacking energy model is very different from that of 
the binding energy model. Here, we follow the same approach and compute the 
averaged fraction of stacking loops of type ${AA \atop UU}$ (or ${UU \atop 
AA}$) and ${AU \atop AU}$ (or ${UA \atop UA}$) as a function of 
temperature under the constraint, 
\begin{equation}
\frac{1}{N}L \partial_L \ln(Z_a(N;s_1,s_2,h=1,L))|_{L=Lc} = \frac{1}{2}. 
\end{equation}

Similarly, in order to avoid the trivial finite size effects for the quenched 
numerical estimate, we fix the number of AA, AU, UA, UU neighbors in the 
randomly chosen sequences to be 320 each~\cite{altsch85}. 

\begin{figure}[h]
\includegraphics[width=8cm]{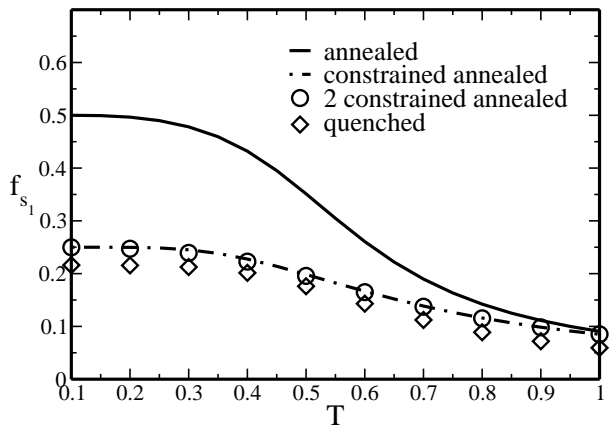}
\caption{Fraction of stacking loops ${AA\atop UU}$ (or ${UU\atop AA}$) in 
the stacking energy model. The constraint of random nearest neighbors fixes 
this quantity much better than averaged number of pairs in the binding energy 
model. The phenomenological
constraint, i.e., a fixed fraction of hairpins, brings this quantity only a 
bit closer to the quenched result.}
\label{frs1}
\end{figure}

\begin{figure}[h]
\includegraphics[width=8cm]{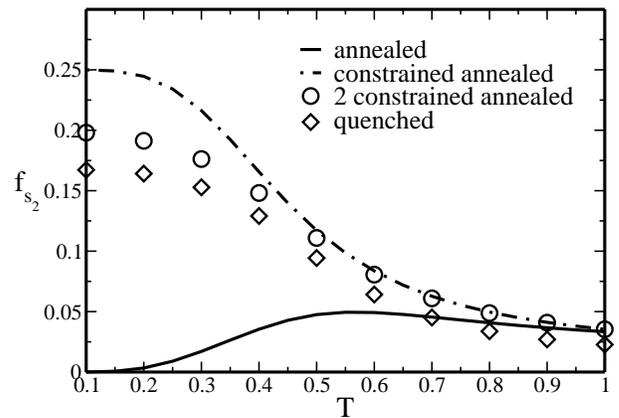}
\caption{Fraction of stacking loops ${AU\atop UA}$ (or ${UA\atop AU}$) in the 
stacking energy model. Again, the constraint of random nearest neighbors 
greatly improves the result. However, unlike the case in Fig.~\ref{frs1}, 
the constraint of a fixed fraction of hairpins also contributes in 
bringing the annealed quantity closer to the quenched result.}
\label{frs2}
\end{figure}

From Figs.~\ref{frs1} and ~\ref{frs2}, we see that the constrained 
annealed results are greatly improved over the plain annealed results. 
This 
verifies the idea that larger deviations from the random disorder result in  
a better correction via the constraint of the random disorder.
For this stacking energy model, at T=0.1, $L_c$=0.0067 is much more different 
from 1 than in the binding energy model. 

From these results, we can see that the constrained annealed ensemble of the 
stacking energy model behaves in the following way. Since the ensemble is 
forced to have the same number of AA (or UU) and AU (or UA) neighbors, 
at zero temperature, the dominating structure is still 
a long stem structure, but with half the stacking loops being of type $s_1$ 
and the other half being $s_2$. This is consistent with 
the fact that fraction of hairpins going to zero as temperature goes to zero 
for the constrained annealed system as shown in Fig.~\ref{frhs12} .

One difference between the quenched ensemble and the constrained annealed 
ensemble is that not all the bases of a random sequence can form stacking 
loops. Thus, we have a finite fraction of hairpins in the quenched ensemble
(Fig.~\ref{frhs12}). 
This difference can used as an additional phenomenological constraint 
to improve the constrained annealed system even further. 

We apply this additional phenomenological constraint by requiring the 
fraction of hairpins $f_h$ to fit the quenched numerical 
estimates and neighboring bases to be uncorrelated at the same time, i.e., to 
enforce 
\begin{small}
\begin{eqnarray}
\frac{L}{N}\partial_L \ln(Z_a(N;s_1,s_2,h,L))|_{h=h_c,L=L_c}\!&\!=\!&\!\frac{1}{2} \\
\frac{h}{N}\partial_h \ln(Z_a(N;s_1,s_2,h,L))|_{h=h_c,L=L_c}\!&\!=\!&\!f_h(T),  
\end{eqnarray}
\end{small}
where $f_h(T)$ is the quenched numerical estimate in this equation. 

From Figs.~\ref{frs1} and \ref{frs2}, we see that this additional 
constraint slightly improve the fraction of stacking loops $s_1$, but 
significantly improves the fraction of stacking loops $s_2$. This can 
be understood since the existence of hairpins introduces AU (or UA) 
neighbors, if the fraction of AU (or UA) neighbors is also required to 
be one half, it will decrease the fraction of stacking loops $s_2$ 
among the stem structures. 

\section{Conclusion}
 
We conclude that the deviation of the annealed ensemble from the quenched 
ensemble is strongly related to the energy model and can be completely biased 
when the correlation is strongly coupled to the energy of the system. 
Quantifying this deviation allows us to do constrained annealing which 
brings the predictions of thermodynamic quantities much closer to the real 
values in the 
quenched ensemble. As the deviation is larger, the constraint is stronger 
and thus brings the annealed ensemble even closer to the quenched results. 
Unfortunately, the biasing toward the quenched ensemble is not strong 
enough to actually drive the system into the glass transition. 

Besides the nearest neighbor correlations, one could also 
consider the correlations for next nearest neighbors or even two bases 
separated by arbitrary distances. In principle, all these correlations 
together 
would bring us to the exact quenched results and thus to the glass 
transition. However, the calculations become much more cumbersome as 
one increases the distance between the two bases, and are left for future 
work. 

\section{Appendix}

\subsection{Annealed partition function for the binding energy model}

The annealed partition function is obtained by first summing over all 
compatible sequences given a secondary structure {\bf S} and then summing 
over all possible structures {\bf S}, which can be done via the recursive 
relation in Fig.~\ref{molten}. 
We define the annealed partition function for a sequence of length N as 
$Z_a(N+1)$. In addition, the annealed partition function for a 
sequence of length N with its two end bases paired is defined as $A_e(N-1)$. 
The recursive relation in 
Fig.~\ref{molten} is then read as 
\begin{equation}
Z_a(N+1)=\frac{1+L}{2}Z_a(N)+\sum_{k=1}^{N-1}\frac{1+L}{4}Z_a(k-1)A_e(N-k).
\end{equation}
The factor $(1+L)/2$ for the first term on the right hand side comes from 
the contribution in nearest neighbor correlations between the free base N 
and base N-1, and the 2 takes care of averaging over the number of sequences.
We have a similar factor in the second term coming from the 
correlation between base k of the arch and base k-1. In the 
later part we will show that the behavior of the annealed partition function 
is mainly determined by the arch term $A_e$, so we will only look at this 
quantity here. The first base of $A_e$ is also specified to be A and 
the last base to be the conjugate base U. 

\begin{figure}[h]
\includegraphics[width=8cm]{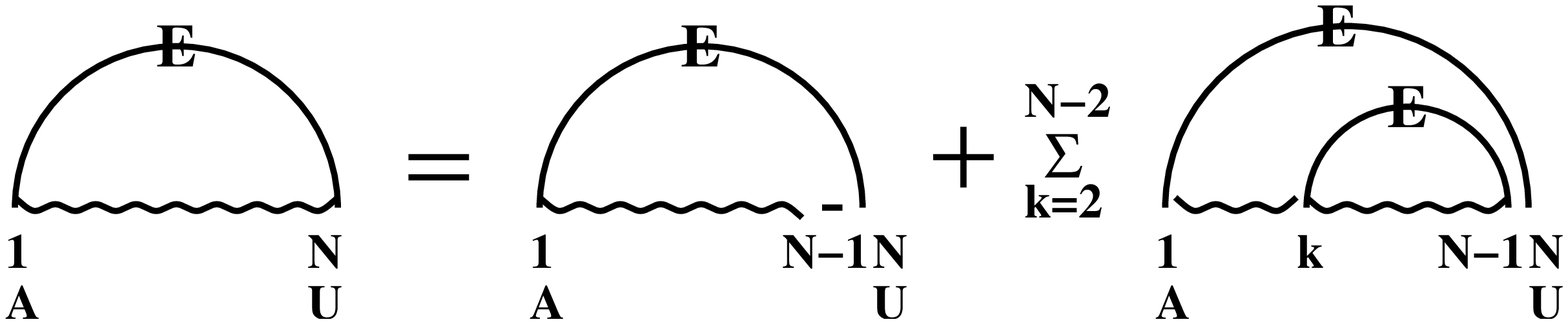}
\caption{Recursive relation for the annealed partition function over 
heterogeneous sequences where the first and the last bases form a conjugate 
pair. A letter 'E' is used to denote that the two bases at the ends of the 
arch are conjugate bases.}
\label{Aq}
\end{figure}

\begin{figure}[h]
\includegraphics[width=8cm]{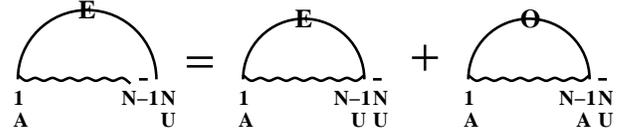}
\caption{Decomposition of an arch with its last base inside being a free 
base, which can be either A or U, into two cases. The letter 'O' is used 
to denote that the two bases at the ends of the arch are non-conjugate.}
\label{Aq1}
\end{figure}

\begin{figure}[h]
\includegraphics[width=8cm]{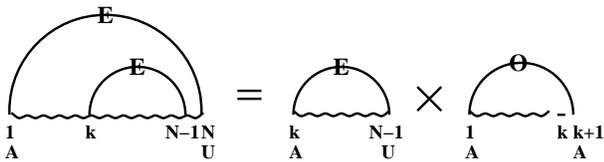}
\caption{Separation of arches.} 
\label{Aq2}
\end{figure}

Again, the annealed partition function for the arch can be obtained through 
a similar recursive relation (Fig.~\ref{Aq}). The two terms on the 
right hand side are further decomposed in Figs.~\ref{Aq1} and 
\ref{Aq2}. 

In these relations, we need to keep track of two factors: the energy 
contributions 
and the nearest neighbor correlations. From the energetic point of view, 
an arch can be thought of simply contributing a Boltzmann factor $q$ and 
need not stand for a real binding pair, even though initially it is used to 
represent a real binding pair. Thus, in Fig.~\ref{Aq1}, as we try to 
relate the annealed partition function to its shorter length ones, we 
assume an effective binding pair between bases 1 and N-1 simply to 
conserve the energy contribution. In this case, the two bases are not 
really paired. 

In order to keep track of the correct nearest neighbor correlations, we use a 
letter E on an arch to denote that the two bases at the ends of the arch 
are conjugate bases. Similarly, a letter O is used to represent two 
non-conjugate bases at the ends of the arch. Thus, in Fig.~\ref{Aq1}, 
the two cases where base N-1 is either A or U are separated and are denoted 
by letter E and O, which is determined by whether the bases 1 and N-1 
are conjugate 
or not. These notations enable us to connect the decomposed terms recursively 
back to the relation in Fig.~\ref{Aq}.

In Fig.~\ref{Aq2}, an inner arch can be treated as a free base in 
considering the energy and correlations for the rest of the bases outside the 
inner arch. However, there is a difference in counting neighbor correlations 
for this treatment because the free base looks as a base A from the right, 
but as a base U from the left. The correct correlations can be obtained if 
we shift this discrepancy to the last base and flip it from U to A. Thus, 
the last term carries a letter O on the arch instead of E.  

These recursive relations are then read as
\begin{small}
\begin{eqnarray}
A_e(N-1)&=&\frac{L}{2} A_e(N-2) + \frac{1}{2}A_o(N-2) + \\ \nonumber
&& \frac{1}{4}\sum_{k=2}^{N-2} A_e(N-k-1)\left[L A_o(k-1)+A_e(k-1)\right], \\ 
A_o(N-1)&=&\frac{L}{2}A_o(N-2)+\frac{1}{2}A_e(N-2) + \\ \nonumber
&& \frac{1}{4}\sum_{k=2}^{N-2} A_e(N-k-1)\left[L A_e(k-1)+A_o(k-1) \right]. 
\end{eqnarray}
\end{small}

Together with the initial conditions, $A_e(1)=q$, $A_e(2)=q L$, $A_o(1)=q L$, 
$A_o(2)=q(1+L)/2$, one can solve for $A_e(N)$ by performing the z-transform 
\begin{eqnarray}
\widehat{A_e}(z) &=& \sum_{N=1}^{\infty} A_e(N) z^{-N}, \\
\widehat{A_o}(z) &=& \sum_{N=1}^{\infty} A_o(N) z^{-N},
\end{eqnarray}
on the recursive relations. After solving for $\widehat{A_e}(z)$, $A_e(N)$ can 
be obtained through the inverse transform
\begin{equation}
A_e(N)=\frac{1}{2\pi i}\oint \widehat{A_e}(z) z^{N-1} dz. 
\end{equation}
From previous studies~\cite{bund02}, we know that in the thermodynamic limit, 
the partition function has an analytical form as 
$A_e(N)\propto N^{-3/2}z_c(q,L)^N$, 
where $z_c(q,L)$ is the greatest real part among the branch points obtained 
from the solution of $\widehat{A_e}(z)$. 

Similarly, if we perform z-transform on equation 15, we can relate the 
z-transform of the annealed partition function $\widehat{Z_a}(z)$ to that of 
the arch $\widehat{A_e}(z)$. Since these two share the same branch points, 
the asymptotic 
behavior of the annealed partition function is different from the above 
formula for the arch by just a different prefactor, which does not play a 
role in the thermodynamic limit.

The fraction of AA (or UU) neighboring bases per base of the annealed system 
is then easily calculated as $L \partial_L \ln(z_c(q,L))|_{L=1}$. 
Unfortunately, the analytical solution of this set of polynomial equations 
is too cumbersome to convey any useful 
information. Thus, we resort to numerical evaluation of this analytical 
solution in this paper. 

\subsection{Fraction of minimal hairpins at zero temperature}

\begin{figure}[h]
\includegraphics[width=8cm]{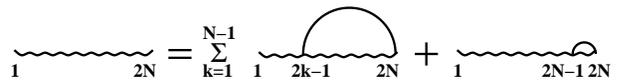}
\caption{Recursive relation for the partition function where all the bases 
are paired.} 
\label{mhairpin}
\end{figure}

As discussed in the main text, the fraction of minimal size hairpins can be 
easily obtained once we figure out the partition function. At zero 
temperature, the partition function is simpler than the finite temperature 
one since we only need to consider the ground states where all bases are 
paired. 
This partition function is obtained through the recursive relation in
a similar way as shown in Fig.~\ref{mhairpin}.

We define the partition function for a sequence of length 2(N-1) as
$Z_m(N,h)$, where $h$ is the Boltzmann factor for a minimal size 
hairpin. The recursive relation is then read as
\begin{equation}
Z_m(N+1)=\sum_{k=1}^{N-1} Z_m(k)Z_m(N-k+1)+h Z_m(N).
\end{equation} 
Together with the initial conditions, $Z_m(1)=1$ and $Z_m(2)=h$,
one can obtain the asymptotic behavior through z-transform. After
simple algebra, we have the largest pole $z_c(h)=h+2\sqrt{h}+1$
for the z-transform of partition function $\widehat{Z_m}(z,h)$.
The partition function $Z_m(N)$ is then proportional to $z_c(h)^N$.
 
The fraction of minimal size hairpins per two bases is then easily calculated 
as
\begin{equation}
\partial_h \ln z_c(h)|_{h=1}=\frac{1+1/\sqrt{h}}{h+2\sqrt{h}+1}|_{h=1}=1/2.
\end{equation} 
Thus, the fraction of minimal size hairpins per base is 1/4.

\subsection{Annealed partition function for the stacking energy model}

\begin{figure}[h]
\includegraphics[width=8cm]{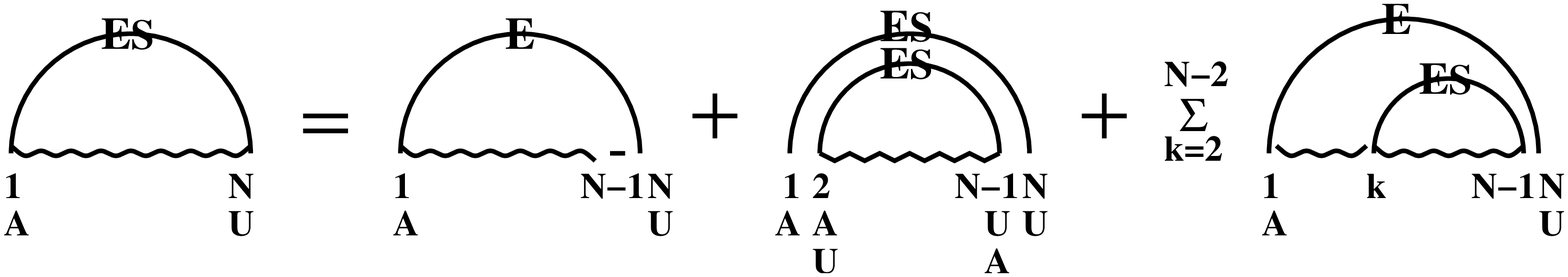}
\caption{Recursive relation for the stacking energy model.} 
\label{Ash}
\end{figure}

The calculation for the stacking energy model follows the same approache. 
However, it is a bit more complicated since we need to keep track of 
stacking loops involving four bases which leads us to the recursive relation 
depicted in Fig.~\ref{Ash}. 

In these recursive relations, we use an additional letter S on the arch to 
denote the fact that we consider the stacking energy of the stacking loop 
formed partly by that binding pair. 
Independent of the type of the arch, all the stacking energies 
inside the arches are still considered in all cases. Thus, the first term 
on the right hand side in Fig.~\ref{Ash} does not contain an S because its 
base N-1 is unbound, and no stacking loop can be formed with the binding 
pair of the arch. 

\begin{figure}[h]
\includegraphics[width=8cm]{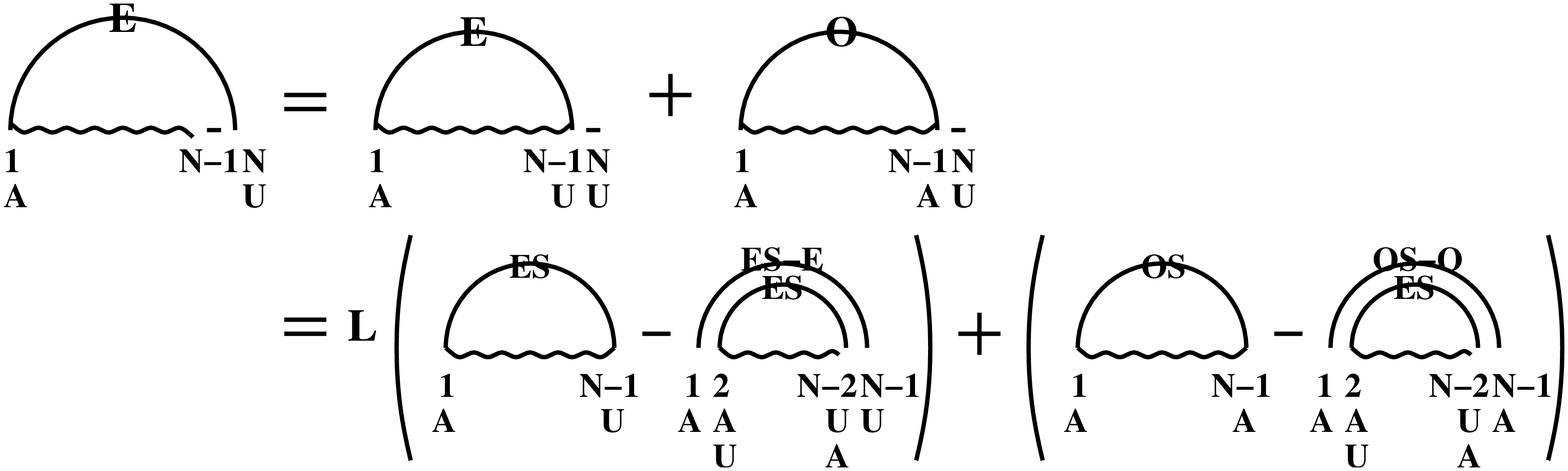}
\caption{Decomposition of the annealed partition function which last base 
inside the arch is a free base.} 
\label{Ash1}
\end{figure}

Similar to the recursive relation in previous section, we then discard the 
last base as a free base as shown in Fig.~\ref{Ash1}. Again, the arches on the right 
hand side are meant to preserve the energy contributions only. In the second 
line of the relation, we further decompose the terms in order to relate these 
terms with the first recursive relation in Fig.~\ref{Ash}. 

\begin{figure}[h]
\includegraphics[width=8cm]{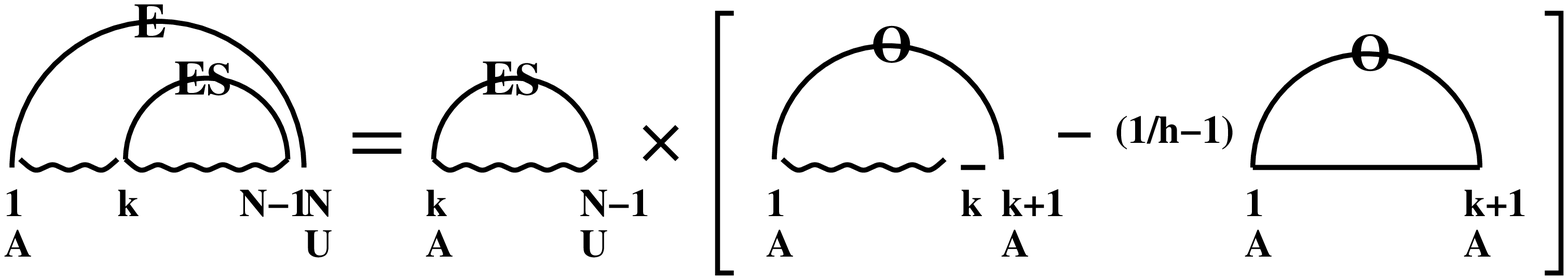}
\caption{Separation of arches considering the hairpin contribution.} 
\label{Ash2}
\end{figure}

In Fig.~\ref{Ash2}, we also separate the contributions of the inner arch 
from the rest part as in Fig.~\ref{Aq2}. One difference is that we consider 
the contribution from the hairpin loops in this case. Thus, the hairpin loop 
contained in the outer arch term is not a real hairpin loop because of the 
existence of the inner arch. The correct result is obtained by adding the 
last term in the relation.

In this stacking energy model, we denote the annealed partition function for 
an arch of length N-1 as $A_{es}(N)$. The recursive relations are then 
read as follows
\begin{widetext}
\begin{footnotesize}
\begin{eqnarray}
A_{es}(N\!-\!1)\!&=&\!\frac{L}{2}\left(A_{es}(N\!-\!2)\!-\!(s_1\!-\!1)\frac{(1+L^2)}{4}A_{es}(N\!-\!4)\right)+\frac{1}{2}\left(A_{os}(N\!-\!2)\!-\!(s_2\!-\!1)\frac{2L}{4}A_{es}(N\!-\!4)\right)+\frac{s_1 L^2+s_2}{4}A_{es}(N\!-\!3) \\ \nonumber
\!&+&\! \frac{1}{4}\sum_{k=3}^{N\!-\!2} A_{es}(N\!-\!k\!-\!1) \left[ L\left(A_{os}(k\!-\!1)\!-\!(s_1\!-\!1)\frac{2L}{4}A_{es}(k\!-\!3)\right)+\left(A_{es}(k\!-\!1)\!-\!(s_2\!-\!1)\frac{1+L^2}{4}A_{es}(k\!-\!3)\right)+2(\frac{1}{h}\!-\!1)H_o(k) \right], \\
A_{os}(N\!-\!1)\!&=&\!\frac{L}{2}\left(A_{os}(N\!-\!2)\!-\!(s_1\!-\!1)\frac{2L}{4}A_{es}(N\!-\!4)\right)+\frac{1}{2}\left(A_{es}(N\!-\!2)\!-\!(s_2\!-\!1)\frac{1+L^2}{4}A_{es}(N\!-\!4)\right)+\frac{s_1 L+s_2 L}{4}A_{es}(N\!-\!3) \\ \nonumber
\!&+&\!\frac{1}{4}\sum_{k=3}^{N\!-\!2} A_{es}(N\!-\!k\!-\!1) \left[ L\left(A_{es}(k\!-\!1)\!-\!(s_1\!-\!1)\frac{1+L^2}{4}A_{es}(k\!-\!3)\right)+\left(A_{os}(k\!-\!1)\!-\!(s_2\!-\!1)\frac{2L}{4}A_{es}(k\!-\!3)\right)+2(\frac{1}{h}\!-\!1)H_e(k) \right],
\end{eqnarray}
\end{footnotesize}
\end{widetext}
where the terms $H_e$ and $H_o$ stand for the contribution from a hairpin. 
They are obtained separately from a recursive relation similar to the one 
in Fig.~\ref{Aq1}, by just replacing the wavy line by a straight line, 
which 
means that bases are not bound. One can then easily formulate the 
recursive relations for $H_e$ and $H_o$.

Together with the initial conditions: $A_{es}(1)=h$, $A_{es}(2)=h L$, 
$A_{es}(3)=h(1+3L^2+s_2+s_1 L^2)/4$, $A_{os}(1)=h L$, $A_{os}(2)=h(1+L^2)/2$, 
$A_{os}(3)=h(3L+L^3+s_1 L+s_2 L)/4$, we can perform z-transform to obtain the 
asymptotic behavior of the annealed partition function for the stacking energy 
model.


\begin{thebibliography}{99}

\bibitem{dill95}
K. A. Dill \textit{et al.},  Protein Sci. {\bf 4}, 561 (1995).

\bibitem{onuch97}
J. N. Onuchic, Z. Luthey-Schulten, and P. G. Wolynes, Annu. Rev. Phys. Chem. 
{\bf 48}, 545 (1997).

\bibitem{garel97}
T. Garel, H. Orland and E. Pitard, J. Phys. I (France) {\bf 7}, 1201 (1997).

\bibitem{shakh97}
E. I. Shakhnovich, Curr. Opin. Struct. Biol. {\bf 7}, 29 (1997).

\bibitem{higgs00}
P. G. Higgs, Q. Rev. BioPhys. {\bf 33}, 199 (2000).  

\bibitem{higgs96}
P. G. Higgs, Phys. Rev. Lett. {\bf 76}, 704 (1996).

\bibitem{pagna00}
A. Pagnani, G. Parisi and F. Ricci-Tersenghi, Phys. Rev. Lett. {\bf 84}, 2026 
(2000).

\bibitem{hart01}
A. K. Hartmann, Phys. Rev. Lett. {\bf 86}, 1382 (2001).

\bibitem{pagna01}
A. Pagnani, G. Parisi and F. Ricci-Tersenghi, Phys. Rev. Lett. {\bf 86}, 1383 
(2001).

\bibitem{bund02}
R. Bundschuh and T. Hwa, Phys. Rev. E {\bf 65}, 031903 (2002).

\bibitem{krzak02}
F. Krzakala, M. M\'ezard and M. M\"uller, EuroPhys. Lett. {\bf 57}, 752 (2002)

\bibitem{marin02}
E. Marinari, A. Pagnani and F. Ricci-Tersenghi, Phys. Rev. E. {\bf 65}, 
041919 (2002)

\bibitem{muller02}
M. M\"uller, F. Krzakala and M. M\'ezard  Euro. Phys. J. E. {\bf 9}, 67 (2002).

\bibitem{orland02}
H. Orland, and A. Zee, Nucl. Phys. B. {\bf 620}, 456 (2002)

\bibitem{mukho03}
R. Mukhopadhyay, E. Emberly, C. Tang and N. S. Wingreen, Phys. Rev. E. {\bf 68}, 041904 (2003)

\bibitem{baiesi03}
M. Baiesi, E. Orlandini and A. L. Stella, Phys. Rev. Lett. {\bf 91}, 198102 
(2003)

\bibitem{leoni03}
P. Leoni and C. Vanderzande, Phys. Rev. E. {\bf 68}, 051904 (2003)
 
\bibitem{tinoco}
I. Tinoco Jr. and C. Bustamante, J. Mol. Biol. {\bf 293}, 271 (1999), and 
references therein.

\bibitem{degenn68}
P. G. de Gennes, Biopolymers {\bf 6}, 175 (1968).

\bibitem{water78}
M. S. Waterman, {\em Advances in Mathematics, Supplementary studies}, 
edited by G.-C. Rota (Academic, New York, 1978), pp.167-212. 

\bibitem{mccas90}
J. S. McCaskill, Biopolymers {\bf 29}, 1105 (1990)  

\bibitem{zuker84}
M. Zuker and D. Sankoff, Bull. Math. Biol. {\bf 46}, 591 (1984)

\bibitem{morita64}
T. Morita, J. Math. Phys. {\bf 5}, 1401 (1964)

\bibitem{orlandi02}
E. Orlandini, A. Rechnitzer and S. G. Whittington, J. Phys. A. {\bf 35}, 7729 
(2002)

\bibitem{altsch85}
S. F. Altschul and B. W. Erickson, Mol. Biol. Evol. {\bf 2}, 526 (1985)


\end{thebibliography}
\end{document}